\documentclass[12pt,twoside]{article}
\usepackage{cite}
\usepackage{psfig}
\usepackage{fancyheadings}
\advance \headheight by 3.0truept       

\renewcommand{\thefootnote}{\fnsymbol{footnote}}

\newcommand{\ov}{\overline}

\def\openone{\leavevmode\hbox{\small1\kern-3.8pt\normalsize1}}%

\newcommand{\eq}[1]{Eq.~(\ref{#1})}
\newcommand{\fig}[1]{Fig.~\ref{#1}}

\newcommand{\tb}[1][]{\ensuremath{\tan^{#1}\!\beta}}
\newcommand{\aS}[1][]{\ensuremath{\alpha_s^{#1}}}
\newcommand{\mg}[1][]{\ensuremath{M_{\tilde{g}}^{#1}}}
\newcommand{\msb}[2][]{\ensuremath{m_{\tilde{b}_{{#2}}}^{{#1}}}}
\newcommand{\mst}[2][]{\ensuremath{m_{\tilde{t}_{{#2}}}^{{#1}}}}
\newcommand{\BRbsg}{\ensuremath{{\cal BR}(b\to s\gamma)}}

\newlength{\nseparation}
\begin{document}
\onecolumn
\begin{titlepage}
\begin{tabular}{l}
Fermilab--Pub--00/242--T\\
CERN-TH/2000-295\\
SCIPP 00/34
\end{tabular}
\hfill
\begin{tabular}{l}
ANL-HEP-PR-00-104 \\
EFI-2000-35 \\
hep-ph/0010003
\end{tabular}
\vskip1truecm
\begin{center}
\boldmath
{\Large \bf $b \to s \gamma$ and supersymmetry with large
$\tan \beta$ }
\unboldmath

\vspace{1cm}
{\sc Marcela Carena}${}^{1,4,}$\footnote{E-mail: carena@fnal.gov},
{\sc David Garcia}${}^{2,}$\footnote{E-mail: David.Garcia.Muchart@cern.ch}, \\
{\sc Ulrich Nierste}${}^{1,}$\footnote{E-mail: nierste@fnal.gov, 
address after 1 Oct 2000: CERN, Theory Division, 1211 Geneva 23, 
                           Switzerland.}
and
{\sc Carlos E.M. Wagner}${}^{3,4,}$\footnote{E-mail: cwagner@hep.anl.gov}
\\[0.5cm]
\vspace*{0.1cm} ${}^1${\it Fermi National Accelerator Laboratory, 
                           Batavia, IL 60510-500, USA.
                      }\\
\vspace*{0.1cm} ${}^2${\it CERN, Theory Division, CH-1211 Geneva 23, 
                           Switzerland.
                      }\\
\vspace*{0.1cm} ${}^3${\it Argonne, Theoretical HEP Division, 
                        Argonne, IL 60439, USA, and\\
Enrico Fermi Institute, Univ.~of Chicago, 5640 Ellis Ave, Chicago, IL
                60637, USA.
                        }\\
\vspace*{0.1cm} ${}^4${\it Santa Cruz Institute of Particle Physics, \\ 
University of California, Santa Cruz,
CA 95064, USA}\\


\vspace*{1truecm}

{\large\bf Abstract\\[10pt]} 
\parbox[t]{\textwidth}{ 
We discuss the constraints on the
supersymmetric parameter space 
from the decay mode $b \to s \gamma$
for large values of $\tan \beta$. We improve
the
theoretical prediction for the decay rate 
by summing
very large radiative corrections to all orders
in perturbation theory. This extends the validity of the perturbative
calculation to the large $\tan\beta$ regime.  This resummation 
of terms of order $\aS[n] \tan^{n+1}\!\beta$ is based on a 
recently proposed effective lagrangian 
for the Yukawa interaction of bottom quarks.
Moreover, we identify an additional source of $\tan\beta$-enhanced
terms, which are of order  $\aS \tan \beta $ and
involve the charged Higgs boson, and analyse their behaviour in higher 
orders of perturbation theory.
After correcting the current  expressions for this rare decay
branching ratio at next-to-leading order,
we obtain that, contrary to recent claims, the measured 
branching ratio of $b \to s \gamma$ constrains the supersymmetric
parameter space in a relevant way, even if $\tan \beta$ is large. 
}
\end{center}

\end{titlepage}

\thispagestyle{empty}
\vbox{}
\newpage
\setcounter{page}{1}
\setcounter{footnote}{0}
\renewcommand{\thefootnote}{\arabic{footnote}}

The measured branching ratio \BRbsg\ is known to provide a valuable
constraint on the parameter space of the Minimal Supersymmetric
Standard Model (MSSM). Increasing attention is devoted to scenarios of
the MSSM with a large value of $\tan \beta$, the ratio of the two
Higgs vacuum expectation values: this region of the parameter space is
experimentally least constrained by the bounds coming from Higgs
searches at the LEP experiments~\cite{lep}.  Theoretical interest in
large $\tan \beta$ scenarios stems from GUT theories with bottom--top
Yukawa unification, which require $\tan \beta = {\cal O} (50)$
\cite{gut,gut2,gut3}.  Recently it has been claimed~\cite{db} that, if
$\tan\beta$ is sufficiently large, the next-to-leading-order
corrections to \BRbsg\ \cite{cdgg,cdgg2} can wash out the constraints
on the MSSM parameter space which arise from the leading-order
calculation~\cite{bbmr}.  This claim is based on the fact that, at
next-to-leading order, two large $\tan\beta$-enhanced corrections
occur: first, supersymmetric QCD (SQCD) corrections to the
chargino-mediated transition lead to terms proportional to $\aS
\tan^2\!\beta$.  Secondly, new $\tan\beta$-enhanced contributions,
absent at leading order, appear in SQCD corrections to the charged
Higgs diagrams. The latter terms are of order $\aS \tan\beta$.
Therefore, for sufficiently large values of $\tan\beta$ the
next-to-leading-order corrections may be of the same order as the
leading-order ones, which involve one power of $\tan\beta$ less.
Cancellations between the leading-order and the next-to-leading-order
contributions of the supersymmetric particles to the decay amplitude
may occur, depending on the specific values of the
supersymmetry-breaking parameters.  It was claimed in Ref.~\cite{db}
that such cancellations do occur in the minimal supergravity model.
In this letter, we shall analyse the dominant, $\tan\beta$-enhanced,
corrections to \BRbsg. We shall explain how to resum large radiative
corrections to this rare decay branching ratio to all orders in
perturbation theory. After this, we shall reanalyse the modifications
of the bounds on the minimal supergravity model parameter space which
arise after radiative corrections are included. We conclude that the
restriction on the sign of the Higgsino mass parameter $\mu$, obtained
at leading order~\cite{gut3}, is robust under the inclusion of
higher-order corrections.

In a previous
publication \cite{cgnw} we have analysed the $\tan \beta$ -enhanced
radiative corrections, which stem from the renormalization of the
Yukawa coupling to down-type fermions. These corrections can
efficiently be summed to all orders in perturbation theory and can be
cast into an effective lagrangian:
\begin{eqnarray}
\label{effL}
 {\cal L}  = - h_d^{ij} \bar{d}_R^i H_1 Q_L^j 
        - \delta h_d^{ij} \bar{d}_R^i H_2 Q_L^j     
 - \hspace{-1ex}h_u^{ij} \bar{u}_R^i (i \tau_2 H_2^*) Q_L^j  
         - \delta h_u^{ij} \bar{u}_R^i (i \tau_2 H_1^*) Q_L^j  
         + \mathrm{h.c.},
\end{eqnarray}
where $\tau_2$ is the usual two by two Pauli matrix, $Q_L = (u,d)_L$,
and a gauge-invariant contraction of weak and colour indices has been
implicitly assumed. In the above, we have ignored small $SU(2)_L$
breaking effects. The dominant contributions to the couplings $\delta
h_d$ and $\delta h_u$ are induced via SQCD corrections. Assuming that
the right- and left-handed soft-supersymmetry-breaking mass parameters
are generation-independent, they are proportional to the couplings
$h_d$ and $h_u$:
\begin{eqnarray}  
\delta h_d^{\mathrm{SQCD}} &  = &  h_d \;
\frac{2\aS}{3\pi}\, \mg 
                        \mu\, \,I(\msb{L},\msb{R},\mg)\, ,
\nonumber\\
\delta h_u^{\mathrm{SQCD}} & = & h_u \;
\frac{2\aS}{3\pi}\, \mg 
                        \mu\, \,I(\mst{L},\mst{R},\mg)\, ,
\label{Dhbt}
\end{eqnarray}
where \mg\ is the gluino mass and $\msb{L,R}$, $\mst{L,R}$ are the
left- and right-handed mass parameters of the down- and up-squarks
respectively.  The dependence of this loop integral on its parameters
is given by
\begin{eqnarray}
I(a,b,c) \; = \; \frac{1}{(a^2-b^2)(b^2-c^2)(a^2-c^2)} 
 \left(a^2b^2\log{\frac{a^2}{b^2}}
        +b^2c^2\log{\frac{b^2}{c^2}}
        +c^2a^2\log{\frac{c^2}{a^2}}\right) . 
\end{eqnarray}
Once SU(2) breaking effects are included, these left- and right-handed
squark mass parameters should be replaced by the squark mass
eigenvalues, \mst{1,2}, \msb{1,2} (or eventually $m_{\tilde{s}_L}$ in
the charged Higgs boson vertex corrections involving the left-handed
strange quark)~\footnote{The vertex corrections to the charged and
neutral Higgs will involve the superpartners of the corresponding
quarks appearing in the external legs.}.  Also, the appropriate CKM
angles, which distinguish the charged and neutral Higgs couplings,
should be included. A precise description of the $\tan \beta$-enhanced
couplings, proportional to $h_b$, of the Higgs particles to top and
bottom quarks was discussed in detail in Ref.~\cite{cgnw}.  In our
application to $b \to s \gamma$ we need the coupling of the charged
Higgs boson to the right-handed bottom and left-handed top quarks, for
which, ignoring small CKM angle effects, an all-order resummation of
the large $\tan\beta$-enhanced corrections is achieved by replacing
the tree-level relation between the coupling $h_b$ in Eq.~(\ref{effL})
and the bottom mass by
\begin{eqnarray}
h_b &=& \frac{g}{\sqrt{2} M_W \cos \beta } 
        \frac{\ov{m}_b (Q)}{1+\Delta m_b^{\mathrm{SQCD}}} \label{hb} .  
\end{eqnarray}
Here $g$ is the SU(2) gauge coupling, $M_W$ is the W-boson mass and 
$\ov{m}_b$ is the bottom mass renormalized in a mass-independent
renormalization scheme like $\ov{\rm MS}$ \cite{bbdm} (as we 
explained in Ref.~\cite{cgnw}, the above relation can
be simply understood in terms of the effective lagrangian of
Eq.~(\ref{effL})). 
The $\ov{t}bH^+$
vertex is renormalized at the scale $Q$, which enters $\ov{m}_b$ in
\eq{hb}. When applied to $b \to s \gamma$ the scale $Q$ equals the 
scale $\mu_{W}$, at which top and $H^+$ are integrated out. 
Following Eqs.~(\ref{effL}) and (\ref{Dhbt}), the
resummed $\tan \beta$-enhanced SQCD corrections 
are contained in~\cite{gut2,gut3} 
\begin{equation}  
\Delta m_b^{\mathrm{SQCD}}  =  \frac{2\aS}{3\pi}\, \mg 
                        \mu\, \tb\,I(\msb{1},\msb{2},\mg)\,
        , \label{dmb}
\end{equation}
where $\aS$ 
should be evaluated at a scale of the order of the masses entering
$I$. If these masses differ so much that they induce large logarithms, 
the scale of $\aS$ is set by the largest of these
masses. The necessity of including these large
corrections, proportional to $\Delta m_b^{\mathrm{SQCD}}$, 
in the computation of \BRbsg\
at large values of $\tan\beta$, was first emphasized
in Ref.~\cite{Uri}.

To consider the dominant chargino--squark contributions, we can write
an effective lagrangian analogous to Eq.~(\ref{effL}), by ignoring the
corrections $\delta h_{u,d}$ and replacing the charged Higgs and one
of the two quarks by their superpartners. Thereafter, we can proceed
in exactly the same way: by replacing the Yukawa coupling $h_b$ in the
stop--bottom--chargino coupling with \eq{hb} we encounter all terms of
order $\aS[n] \tb[n]$ for $n=0,1,2,3,\ldots$ Since the leading-order
chargino--stop contribution grows linearly with $\tan\beta$, this
resummation leads to contributions to \BRbsg\ of order $\aS[n]
\tb[n+1]$.  Similar to the charged Higgs case, one can therefore
implement the resummation into an existing leading-order calculation
by simply replacing $h_b$ by its expression given in Eq.~(\ref{hb}).
We have done this with the Wilson coefficients $C_7$ and $C_8$, which
contain the supersymmetric terms relevant to the $b \to s \gamma$
amplitude.  After expanding \eq{hb} to first order in \aS\ we have
indeed reproduced the dominant terms of the known
next-to-leading-order result of \cite{cdgg}: these terms of order $\aS
\tb[2]$ stem from the chargino--squark contributions and equal
$-\Delta m_b^{\mathrm{SQCD}}$ times the corresponding leading-order
piece involving $h_b$. While the authors of \cite{db} have correctly
connected the $\aS \tb[2]$ term to the sbottom mixing, they have
erroneously claimed the absence of terms of order $\aS[2] \tb[3]$.
Since the sbottom mixing terms proliferate into the Yukawa counterterm
through $\Delta m_b^{\mathrm{SQCD}}$, they iteratively show up in any
order of perturbation theory, as explained in detail in \cite{cgnw}.
Expanding \eq{hb} to first order in \aS\ also reproduces one term of
order $\aS \tb $ associated with SQCD corrections to the $H^+
\ov{t}_Lb_R$ vertex in the charged Higgs diagram.

Interestingly enough, as has been shown in Ref.~\cite{cdgg},
there is an additional source of $\tan\beta$-enhanced corrections
in the charged Higgs diagrams:
while the tree-level $H^+ \ov{t}_R s_L$ vertex is suppressed by 
$1/\tb$, this vertex
suppression is lifted at the one-loop level, so that the
next-to-leading-order charged-Higgs contribution 
to \BRbsg\ is \tb-enhanced with
respect to the leading-order one.  
This feature originates from the loop-induced
flavour-violating couplings $\delta h_u^{ij}$ described in Eq.~(\ref{effL}),
which involve a right-handed top
quark, a left-handed 
stran\-ge quark, and  $H_1^+$, the charged component of
the doublet $H_1$. 
The disappearance of the $\tan\beta$ suppression is due to the fact
that, at large values of $\tan\beta$, the physical charged Higgs can
be approximately identified with $H_1^+$, while its component on
$H_2^+$ is suppressed by $1/\tan\beta$. The absence of a $\tan\beta$
suppression in this loop-induced coupling of the charged Higgs leads
to a term of order $\aS \tan\beta$ in the charged Higgs diagram, where
the factor of $\tan\beta$ comes from the bottom-quark Yukawa coupling
in Eq.~(\ref{hb}).\footnote{ While the term involving $\delta
  h_u^{ij}$ is important for the $H^+ \ov{t} s$ vertex in the
  effective lagrangian in Eq.~(\ref{effL}), it is subdominant in the
  effective $H^+ \ov{t} b$ coupling.  Therefore it has not been
  considered in Ref.~\cite{cgnw}, which deals with Higgs and top
  decays.}

Schematically, the resummation of the \tb-enhanced corrections
to the chargino and charged Higgs contributions can be summarized
as follows: the $\tan\beta$-enhanced chargino contributions to \BRbsg\ 
are 
\begin{equation}
\left.
\BRbsg \right|_{\chi^{\pm}} \propto 
\mu A_t \tan\beta f(\mst{1},\mst{2},m_{\tilde{\chi}^+})
\frac{m_b}{v (1 + \Delta m_b)} ,
\label{Char}
\end{equation}
where all dominant higher-order contributions are included through 
$\Delta m_b$, and $f$ is the loop inte\-gral appearing at one loop.
The relevant charged-Higgs contributions to \BRbsg\
in the large $\tan\beta$ regime
are 
\begin{equation}
\left.
\BRbsg \right|_{H^+}
\propto \frac{ m_b \left( h_t \cos\beta  - \delta h_t
\sin\beta \right)}{v \cos\beta (1 + \Delta m_b)} \; 
g(m_{H^+},m_t) ,
\label{ChH}
\end{equation}
where we have left the $\cos\beta$ and $\sin\beta$ factors associated
with their sources, and $g$ is the loop-integral appearing at the
one-loop level. In the above $\delta h_t$ proceeds from the
flavour--violating coupling $\delta h_u$ in Eq.~(\ref{effL}), where,
as we already explained, the squark masses should be replaced by the
superpartners of the left-handed strange squark and of the
right-handed top squark.  Since the left-handed and right-handed top
squarks mix in a non-trivial way, the loop integral $I(a,b,c)$ should
be replaced by the sum of two-loop integrals for the two stop mass
eigenstates with appropriate projection factors
$\cos^2\!\theta_{\tilde{t}}$ and $\sin^2\!\theta_{\tilde{t}}$,
respectively \cite{cdgg},
 
\begin{equation}
\delta h_t  = h_t \; \frac{2 \alpha_s}{3 \pi} \; \mu M_{\tilde{g}}
\left( 
\cos^2\!\theta_{\tilde{t}} \; I(m_{\tilde{s}_L},m_{\tilde{t}_2},
M_{\tilde{g}}) 
+
\sin^2\!\theta_{\tilde{t}} \; I(m_{\tilde{s}_L},m_{\tilde{t}_1},
M_{\tilde{g}})  \right) .
\label{deltaht}
\end{equation}

Observe that, as we first explained 
in Ref.~\cite{cgnw} and follows from Eqs.~(\ref{Char}) and (\ref{ChH}), 
the corrections to the bottom-quark Yukawa
coupling originate from mass counterterms and 
lead to terms of order $\tan^n\!\beta$ at $n$th-order of
perturbation theory. The one-loop induced $H^+ \ov{t}_R s_L$
vertex, instead, 
is of order $\tan^0\!\beta$, and higher-order corrections
to it cannot induce any $\tan\beta$-enhanced term. The above
guidelines are useful  to extend the validity of the
calculation presented in Ref.~\cite{cdgg} to large values of
$\tan\beta$.

It should be stressed that, within the conventions of Ref.~\cite{cdgg},
the sign of the one-loop contribution, proportional to  
$\mu H_2(u_i,x_j)$ in the notation of 
Appendix A of Ref.~\cite{cdgg},
to the couplings of the charged
Goldstone and Higgs fields to the right-handed top and left-handed
down-like quark should be the ones that arise from Eq.~(\ref{effL}),
which are opposite to the ones stated in Ref.~\cite{cdgg}.\footnote{
  The authors of Ref.~\cite{cdgg} have independently detected these
  sign errors.  { They posted a
  revised version of Ref.~\cite{cdgg} to the hep-ph archive 
  during completion of this article.}}  
A change of sign in these couplings leads to an
inversion in the sign of those $\tan\beta$-enhanced charged-Higgs
contributions to $b \to s \gamma$ not related with $\Delta m_b$ (the
sign of $\delta h_t$ in Eq.~(\ref{ChH})).  Once the correct sign is
used, we find that the higher-order $\tan\beta$-enhanced corrections
to the charged-Higgs and chargino contributions to $b \to s \gamma$
lead to an enhancement (suppression) of these contributions for
negative (positive) values of $\mu M_{\tilde{g}}$.

At large values of $\tan\beta$, the leading-order char\-gino 
contributions 
to the amplitude of the decay rate $b \to s \gamma$ are
proportional to $A_t \mu$. In supergravity models, the sign
of $A_t$ is opposite to the one of the
gaugino masses. This sign relation holds
unless the boundary values of $A_t$ at
the high-energy input scale is one order of magnitude larger
than the gaugino soft-supersymmetry-breaking mass 
parameters~\cite{gut,iblop}. 
Since the charged Higgs-top
diagram leads to a contribution to the amplitude of
the decay $b \to s \gamma$ of the same sign as the
SM contribution, and the relative sign of the chargino contribution
is governed by the sign of $A_t \mu$, 
negative values of $A_t \mu$
(or equivalently, defining the gaugino masses as positive, positive
values of $\mu$)
are necessary in order to render values of \BRbsg,
in agreement with the ones observed experimentally \cite{bbmr,gut3}. 
The present study shows 
that for negative values
of $\mu$ the next-to-leading-order corrections to the charged
Higgs  and the chargino-stop contributions further enhance the 
$b \to s \gamma$ decay amplitude. Therefore,   
it follows  that even after considering higher-order effects,
positive values of $\mu$ are necessary in order to obtain
correct values for \BRbsg\ within minimal supergravity
models, for which the sign of $A_t$ at low energies tends to be
negative { \cite{dt}}.

Contrary to our results,
by using the expressions given in Ref.~\cite{cdgg}, 
one would obtain
an incorrect strong suppression (and even a change of sign at 
sufficiently large values of
$\tan\beta$) of the charged-Higgs contributions for negative values of
$\mu$. Under these conditions, the authors of
Ref.~\cite{db} incorrectly found that, for negative values of $\mu$,
acceptable values of \BRbsg\ may be obtained.
Analogously, for positive values of $\mu$ they found an incorrect
enhancement of the charged-Higgs contribution, rendering it difficult
to find acceptable values for the rare bottom quark decay rate under
study.

\begin{figure}[t]
\centerline{
\psfig{figure=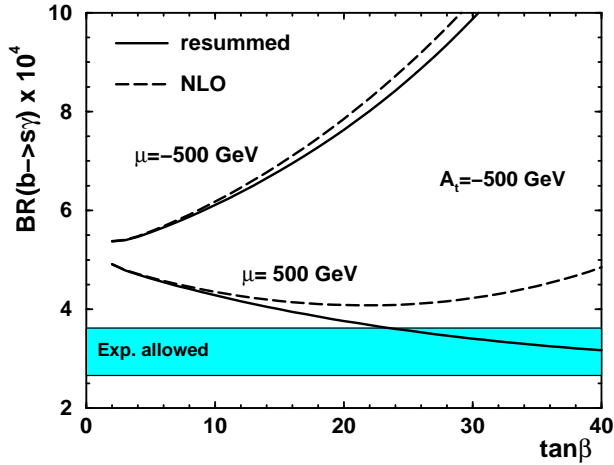,width=8cm}}
\caption{Comparison of the result for ${\cal BR}(b \to s
  \gamma)$, as obtained in this work, as a function of $\tan\beta$, 
with the NLO expression of \cite{cdgg}, where we have replaced
the appropriate signs in the next-to-leading-order
corrections, as  explained in the text.
The charged-Higgs 
  boson mass is $200$~GeV and the light stop mass is $250$~GeV. 
The values of $\mu$ and $A_t$ are indicated in the plot while
  $M_2$, the gluino, heavy-stop and down-squark masses are set at
  $800$~GeV.}
\label{plot1}
\end{figure}

\begin{figure}[t]
\centerline{
\psfig{figure=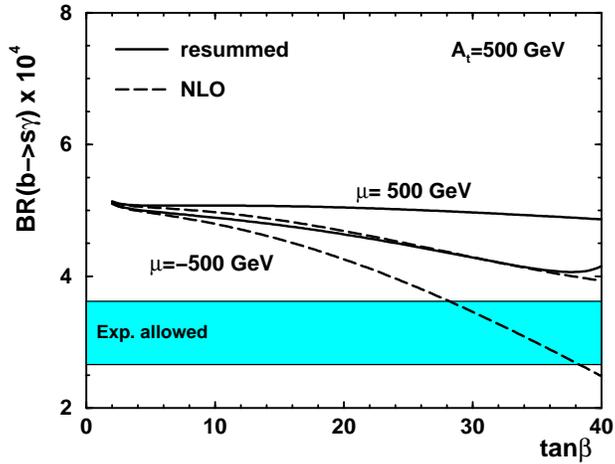,width=8cm}}
\caption{As in Fig.~\ref{plot1}, but for positive $A_t=500$~GeV.}
\label{plot2}
\end{figure}

\begin{figure}[t]
\centerline{
\psfig{figure=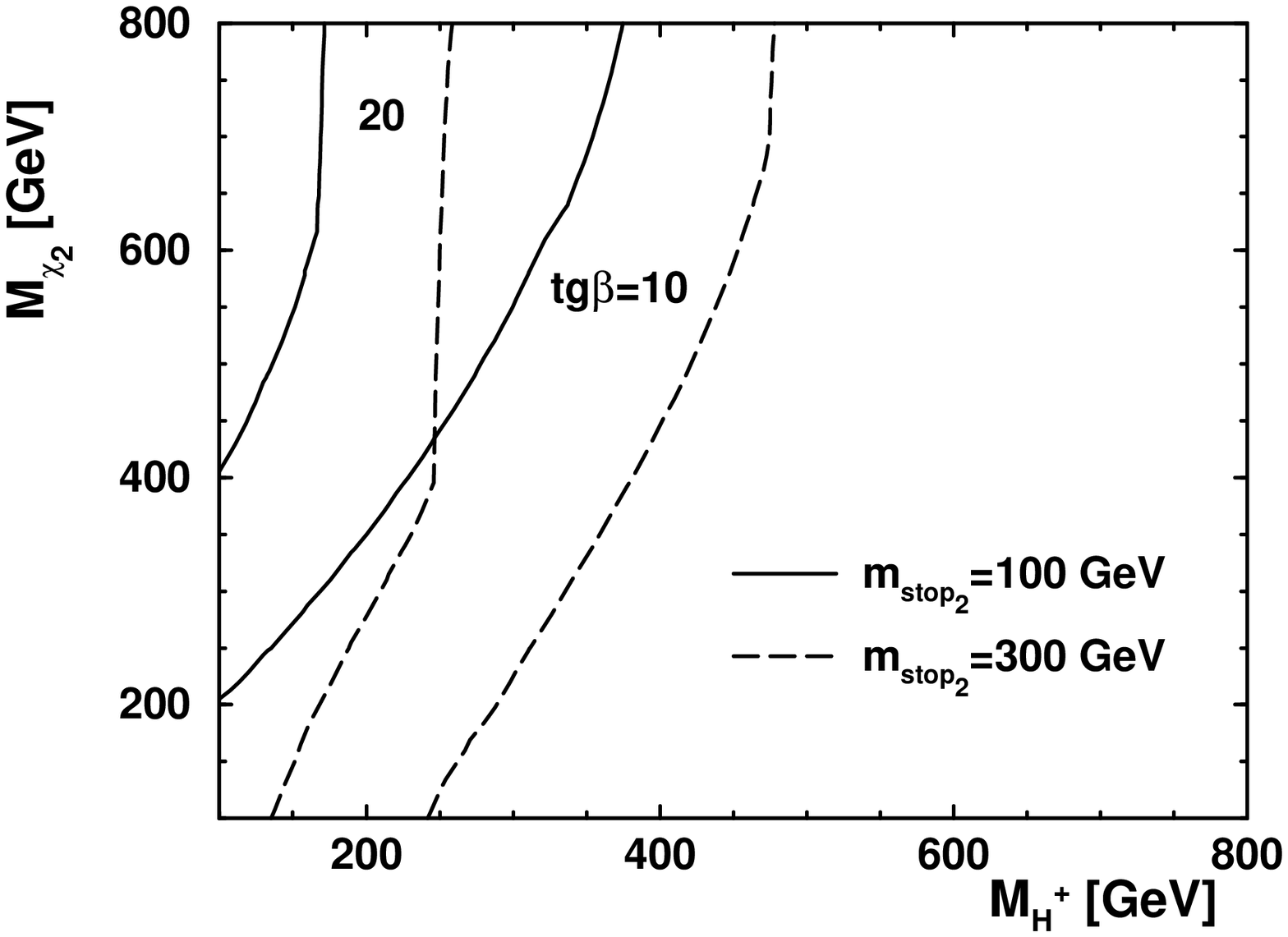,width=8cm}}
\caption{Combined bound on the charged Higgs and chargino masses, for
  various values of the mass of the lightest stop, $m_{\tilde{t}_2}$,
  and \tb. The excluded region corresponds to light (heavy) charged
  Higgs (chargino) masses. We have scanned for
  $m_{\tilde{t}_2}<m_{\tilde{t}_1}\leq 1$~TeV,
  $m_{\tilde{\chi}^+_2}<m_{\tilde{\chi}^+_1}\leq 1$~TeV and $|A_t|\leq
  500$~GeV, the rest of SUSY masses have been set at 1~TeV.}
\label{plot3}
\end{figure}

We give here  a simple recipe, based on Eqs.~(\ref{ChH}) and
(\ref{Char}), for  
implementing the all-order
resummation of $\tan\beta$-enhanced contributions
into existing computations for the $b \to s \gamma$
amplitude: to this end multiply
first in Eq.~(4) of \cite{cdgg} the terms proportional to $1/\cos
\beta$ with $1/(1+\Delta m_b^{\mathrm{SQCD}})$. 
Second, multiply the term proportional to $A_d$ in Eq.~(53) of first
reference in \cite{cdgg2}, which contains the contribution of the
char\-ged Higgs, with $1/(1+\Delta m_b^{\mathrm{SQCD}})$. Third, add a
term of the form $F^{(2)}_i(x)\cdot (1-1/(1+\Delta
m_b^{\mathrm{SQCD}}))$ in Eq.~(28) of first reference in \cite{cdgg2},
for $i=7,8$.
Fourth, delete the term proportional to $\mu \tan \beta H_2(x_1,x_2)$ 
in $H_d$, 
$U_d$ and
$\Delta _d^{(2)}$, which are defined in appendices A.2.--A.4.\ of
\cite{cdgg}.  $\Delta _b^{(2)}$ enters the Wilson coefficients
$C_{7,8}$ through $G_{7,8}^{\chi,2}$ in Eq.~(13) and $H_b$, $U_b$ enter
these coefficients through Eqs.~(25), (26) of \cite{cdgg}.
Multiply the remaining $\tan\beta$-enhanced terms, 
proportional to $\mu \tan\beta H_2(u_i,x_2)$,
in the charged-Goldstone
and charged-Higgs boson contributions by $1/(1+\Delta m_b^{\mathrm{SQCD}})$. 
The presence of these
additional $\tan\beta$-enhanced terms in the charged Goldstone diagram
ensures the proper decoupling of
the supersymmetric corrections in the limit of large supersymmetric
particle masses.

To demonstrate which is the effect of the improvement in the
theoretical prediction for \BRbsg\ developed in this work, we compare
in Figs.~\ref{plot1} and \ref{plot2} our result, including a
resummation of the dominant $\tan\beta$-enhanced radiative corrections
to all orders of perturbation theory, with the next-to-leading-order
result of Ref.~\cite{cdgg}, with the appropriate sign corrections
explained above, for typical values of the supersymmetric parameters.
In \fig{plot1}, a negative value of $A_t=-500$~GeV at low energies is
chosen, as is usually the case in supergravity models. For the
experimental measurement of the $b\to s\gamma$ branching ratio, we use
the combined result of CLEO \cite{CLEO} and ALEPH \cite{ALEPH}, ${\cal
  BR}(b\to s\gamma)=(3.14\pm 0.48)\times 10^{-4}$.  \fig{plot2} shows
the complementary case of positive $A_t=500$~GeV.  Notice that, for
$A_t>0$, the behaviour of the MSSM result never differs crucially from
the SM prediction: there is always some cancellation between competing
terms, either between the charged Higgs and the chargino contributions
(for values of $\mu<0$) or between the LO and the \tb-enhanced NLO
corrections (for values of $\mu>0$).
  
  In \fig{plot3} we confront the resummed theoretical prediction for
  the $b\to s\gamma$ branching ratio with the experimental result to
  find the excluded region --from the curves, to the left-- 
  in the $(M_{H^+}, M_{\chi^+})$ plane. Curves are shown for
  $\tb=10,20$, and $m_{\tilde{t}_2}=100, 300$~GeV, after a scan in the
  remaining parameters, as described in the caption.

In conclusion, we have presented in this article a method allowing
to extend the validity of the NLO computation of the decay
rate $b \to s \gamma$ to the large-$\tan\beta$ regime, which
is based on a resummation to all orders of the dominant
$\tan\beta$-enhanced { radiative} corrections.  We have
proved that, contrary to recent claims, the next-to-leading-order
corrections preserve the basic features of the leading-order
results in constraining positive values of $\mu A_t$ within
minimal supergravity models, for which the low-energy values
of $A_t$ tend to be negative.

\section*{Acknowledgements}
We thank Konstantin Matchev for useful discussions. One of the authors
(DG) thanks the Fermilab Theory Group for their hospitality and
financial support. The work of D.G. was supported by the European
Commission TMR programme under the grant ERBFMBICT 983539.
Work supported
in part by the US Department of Energy, High Energy Physics Division,
under Contracts DE-AC02-76CHO3000 and W-31-109-Eng-38.

\section*{Note added}
{ After completion of this work} similar results have been
presented in \cite{dgg}. While our paper focuses on the all-order
resummation of $\tan \beta$-enhanced SQCD corrections and the
clarification of the findings of \cite{db}, the authors of \cite{dgg}
have { discussed} dominant two-loop contributions arising from
$\tan \beta$-enhanced SQCD and supersymmetric electroweak corrections
and of large logarithms which arise for heavy superpartner masses.

\end{document}